\RequirePackage[l2tabu, orthodox]{nag}
\documentclass[a4paper,11pt]{article}
\usepackage{multicol}
\usepackage{etex}
\usepackage{anysize}

\usepackage[utf8]{inputenc}
\usepackage[english]{babel}

\usepackage[pdftex]{graphicx}

\usepackage{amsmath, amssymb}
\allowdisplaybreaks
\usepackage{mathtools}
\usepackage{nicefrac}
\usepackage{fancyhdr}

\newcommand{\ruleAddHCN}{$r_1$} 
\newcommand{\ruleCNToAmid}{$r_2$} 
\newcommand{\ruleAmidToAcidWater}{$r_3$} 
\newcommand{\ruleWaterToImine}{$r_4$} 
\newcommand{\ruleWaterToImineInverse}{$r_5$} 
\newcommand{\ruleAddAlcohol}{$r_6$} 
\newcommand{\ruleAminal}{$r_7$} 
\newcommand{\ruleAminalInverse}{$r_8$} 
\newcommand{\ruleKetoEnol}{$r_9$} 
\newcommand{\ruleEnolKeto}{$r_{10}$} 
\newcommand{\ruleKetimineEnolimine}{$r_{11}$} 
\newcommand{\ruleEnolimineKetimine}{$r_{12}$} 
\newcommand{\ruleKetoEnolUmpolung}{$r_{13}$} 
\newcommand{\ruleAdditionUmpolung}{$r_{14}$} 
\newcommand{\ruleCNCEnolSwap}{$r_{15}$} 
\newcommand{\ruleAddHCNToAldehyde}{$r_{16}$} 
\newcommand{\ruleEtherBreak}{$r_{17}$} 
\newcommand{\ruleAldolAdditionNInverse}{$r_{18}$} 

\newcommand{\chemfig}[1]{\ensuremath{\mathrm{#1}}}

\newcommand{\removeStuff}[1]{}
\newcounter{incFigCounter}
\stepcounter{incFigCounter}
\newcommand{\incFig}[1]{%
\includegraphics{incFig/\arabic{incFigCounter}.pdf}%
\stepcounter{incFigCounter}%
}

\usepackage{todonotes}
\usepackage{hyperref}
\hypersetup{
} 
\usepackage{datetime}
\usepackage{subfig}
\usepackage[numbers,sort,square]{natbib}
\usepackage{appendix}
\usepackage{moreverb}

\title{\textit{In silico} Support for\\ Eschenmoser's Glyoxylate Scenario}
\author{
\\
Jakob~L.~Andersen$^{1}$,
Christoph Flamm$^{2}$,
Daniel Merkle$^{1}$,
Peter~F.~Stadler$^{2-7}$
\\[1cm]
$~^1$ Department of Mathematics and Computer Science\\ University of Southern Denmark, Denmark\\[0.2cm]
$~^2$ Institute for Theoretical Chemistry\\ University of Vienna, Austria\\[0.2cm]
$~^3$ Department of Computer Science\\ University of Leipzig, Germany\\[0.2cm]
$~^4$ Max Planck Institute for Mathematics in the Sciences\\ Leipzig, Germany\\[0.2cm]
$~^5$ Fraunhofer Institute for Cell Therapy and Immunology\\ Leipzig, Germany\\[0.2cm]
$~^6$ Center for non-coding RNA in Technology and Health\\  University of Copenhagen, Denmark\\[0.2cm]
$~^7$ Santa Fe Institute, Santa Fe, USA
}
\date{}

\marginsize{1.5cm}{1.5cm}{1cm}{1cm}
\usepackage{fancyhdr}
\begin{document}
\maketitle
\begin{abstract}
  A core topic of research in prebiotic chemistry is the search for
  plausible synthetic routes that connect the building blocks of modern
  life such as sugars, nucleotides, amino acids, and lipids to ``molecular
  food sources'' that have likely been abundant on Early Earth. In a recent
  contribution, Albert Eschenmoser emphasised the importance of catalytic
  and autocatalytic cycles in establishing such abiotic synthesis
  pathways. The accumulation of intermediate products furthermore provides
  additional catalysts that allow pathways to change over time. We show
  here that generative models of chemical spaces based on graph grammars
  make it possible to study such phenomena is a systematic manner. In
  addition to reproducing the key steps of Eschenmoser's hypothesis paper,
  we discovered previously unexplored potentially autocatalytic pathways
  from \chemfig{HCN} to glyoxylate. A cascading of autocatalytic cycles
  could efficiently re-route matter, distributed over the combinatorial
  complex network of \chemfig{HCN} hydrolysation chemistry, towards a
  potential primordial metabolism. The generative approach also has it
  intrinsic limitations: the unsupervised expansion of the chemical space
  remains infeasible due to the exponential growth of possible molecules
  and reactions between them. Here in particular the combinatorial
  complexity of the \chemfig{HCN} polymerisation and hydrolysation networks
  forms the computational bottleneck. As a consequence, guidance of the
  computational exploration by chemical experience is indispensable.
\end{abstract}

\section{Introduction}

The Origin of Life is among the most fascinating and most interdisciplinary
scientific problems. Despite a century of research, however, it still
presents itself as an enigma. On the one hand, we still lack both a
detailed understanding of the principles that govern the transition from
non-living matter to living systems in general, and a clear historical
scenario of the emergence of Life on Earth.  See, e.g.,
\citet{Ruiz-mirazo:2014} for a recent review.

A key topic of prebiotic chemistry is to find synthetic routes to the
bio-molecular building blocks of modern life that are chemically plausible
given the environmental conditions on Early Earth. In particular, these
synthetic routes need to start with material that has been reasonable
abundant after the formation of the planet. This is a notoriously difficult
research problem for several reasons: (1) The chemical search space is vast, in
fact it is by far too large to enumerate exhaustively, and it cannot be
confined to only those compounds that are well-characterised and described
in current chemistry databases. (2) The constraints imposed by our current
knowledge of the conditions on early earth are too vague to exclude large
portions of the search space. (3) The problem of finding synthetic routes
to target molecules itself, even for a fixed set of chemical reactions, is
a very difficult combinatorial problem. (4) Experimental verification of
potential routes is inherently slow. Consequently, conceptual guiding by
skilled organic chemists is indispensable to direct research towards
regions of prebiotic chemical space worthwhile of being explored in more
detail. On the other hand, efficient formal, computational approaches are
required to handle the combinatorial complexity in practice.

Recently, Albert Eschenmoser published a conceptual paper \cite{esch07}
detailing the hypothetical relationships between \chemfig{HCN} chemistry
and the constituents of the reductive citric acid cycle. Eschenmoser
suggested to look for catalytic and/or autocatalytic processes in the
non-robust subspace of \chemfig{HCN} chemistry since these have the
potential to canalise this fragile chemistry towards the formation of
\chemfig{C_4} and \chemfig{C_6} compounds, which in turn could function as
precursors of $\alpha$-keto acids and carbohydrates. In this
scenario \cite{Eschenmoser:2007} glyoxylate (CID~760) and its formal dimer,
dihydroxyfumarate (CID~54678503), serve as the pivotal entry points into
sugar chemistry. Both compounds can be formally viewed as
``aquo-oligomeres'' of carbon monoxide (CID~281) and open up a route to
sugars that is independent of formaldehyde (CID~712) as basic building
block. Eschenmoser furthermore pointed out that aldehydes have the
potential to catalyse both the oligomerisation of \chemfig{HCN} (CID~768)
to the tetramer \chemfig{DAMN} (CID~2723951), and the hydrolysation of the
cyanide groups (\chemfig{-CN}) of the \chemfig{DAMN} to
the respective amide groups (\chemfig{-CONH_2}). This 
type of chemistry makes oxaloglycolate (CID~524) (a tautomer of
dihydroxyfumerate) and glyoxylate accessible from \chemfig{HCN}. Finally,
Eschenmoser proposes a hypothetical autocatalytic cycle feeding on
glyoxylate in which oxaloglycolate (or its diamides) acts as
``umpolung-catalyst'' for its on production.
Umpolung is an important
process in organic chemistry where the reactivity of a functional group
(e.g., of a carbonyl group \chemfig{C\!=\!O}) is inverted via chemical
modification. The majority of bonds in organic reactions are formed between
atoms of different polarity. Heteroatoms usually polarise the carbon
skeleton in consequence of their high electronegativity. Therefore the
carbon atom of the carbonyl group is partially positively charged allowing
carbanions (carbon atoms carrying a negative charge) to attack the carbonyl
carbon atom to form a novel carbon-carbon bond. Umpolung now chemically
modifies the carbonyl group in such a way that the polarity of the carbonyl
group is inverted. This means that the carbonyl carbon atom, after
umpolung, carries a negative charge and can, in contrast to its normal mode
of reactivity, attack itself a positively charged carbon center to form a novel
carbon-carbon bond. (for a review on ``umpolung'' in organic synthesis see
e.g., \citet{Seebach:1979})

Albert Eschenmoser arrived at hypotheses put forward in ``The Chemistry of
Life's Origin'' based on his extensive knowledge of organic chemistry. In this
contribution we ask whether these scenarios could also have been found by
formal computational methods and whether there are plausible alternatives,
e.g., other autocatalytic processes hidden in the combinatorial chemical
space of \chemfig{HCN} chemistry. To this end we employ a computational
framework we have developed specifically to explore very large
chemical reaction networks. It combines a generative approach that
implements chemical reactions as graph transformations with network flow
analysis on the resulting hypergraph representing the reaction network \cite{Andersen:13a,Andersen:14a}.
This framework allows, for example, to
detect autocatalytic cycles \cite{Andersen:12a} and to establish the
existence of synthesis routes connecting a pair of compounds \cite{Andersen:14b}.
We will use it here to redraw and elaborate on
Eschenmoser's pictures by computational means.

\section{Formal Approach}

The vastness of chemical spaces renders the common methods of computational
chemistry, from empirical potentials to full-fledged quantum chemistry,
infeasible for large-scale explorations. A purely combinatorial approach
that views molecules as graphs and chemical reactions as graph
transformations, however, brings chemical spaces into reach of present-day
computational capabilities. The use of graphs as models of molecules dates
back to seminal work by Arthur Cayley \cite{Cay1874} and James J.\
Sylvester, who introduced the term ``graph'' in his \emph{Nature} paper
\cite{Syl1878} in 1878 in order to combine algebra and chemistry for the
enumeration of isomers. We follow the same natural paradigm: atoms are the
vertices labelled by atom type and chemical bonds become edges in the
graph, labelled by the bond type. A chemical reaction is then simply a
transformation of a collection of educt graphs into a collection of product
graphs that preserves atom labels. However, not all such transformations
are chemically meaningful. Instead, chemical reactions follow a quite
restrictive set of rules that corresponds to the reaction mechanisms and
``name reactions''.\footnote{\url{http://www.organic-chemistry.org/namedreactions/}}

\subsection{Graph Grammars and the Double Pushout Approach}
\label{sect:gg}

Despite the ubiquity of graphs in the chemistry literature, graph
transformations \cite{Nag79} have been introduced only quite recently as an
explicit models of chemical reactions \cite{Benkoe:03b}. They provide a
more general and more versatile framework for specifying chemical reactions
than earlier, matrix-centred approaches such as the Dugundji-Ugi theory
\cite{Dugundji:73,Fontain:91}.

The basic idea behind graph transformations is that each transformation
rule $r$ specifies that a ``left'' graph $L$ can be replaced by a ``right''
graph $R$. The rule $r$ can be applied to a graph $G$ provided $L$ is
contained as a subgraph in $G$. The application of $r$ to $G$ then consists
in replacing the subgraph $L$ within $G$ by $R$, resulting in a new graph
$H$. A simple example of such a rule is the addition of \chemfig{Br_2} to a
\chemfig{CH_2\!=\!CH_2} double bound resulting in an dibromide
\chemfig{Br\!-\!CH_2\!-\!CH_2\!-\!Br}.  Graph transformation systems generalise the string
or term rewriting systems familiar in mathematics, computer science, and
logic \cite{Baader:08}.

Several different mathematical frameworks have been explored for graph
rewriting \cite{Ehrig:06}. They vary in the exact definitions of how $L$
has to be contained in $G$, i.e., when the precondition for applying a rule
is satisfied, and how exactly the replacement $R$ is inserted into $G$ to
form the rewritten graph $H$. We adopted one of the more restrictive
frameworks known as the Double-Pushout approach (DPO) because it has
several features that are appealing for applications to chemistry. In DPO a
rule, usually called a production, consists of three graphs, the ``educt''
$L$, the ``product'' $R$, and a ``context'' $K$ that remains unchanged
during the reaction. The DPO formalism stipulates that $K$ is contained as
a subgraph in both $L$ and $R$. Since atoms are neither lost nor gained in
chemical reactions, $K$ in particular contains all atoms that take part in
the reaction. Together, $L$, $K$, and $R$ completely specify the formal
transition state and the atom map of the transformation. A further
advantage of DPO is that exchanging the roles of $L$ and $R$ automatically
leads to the production for the reverse reaction. We refer to
\citet{Andersen:13a} for more information on these technicalities.

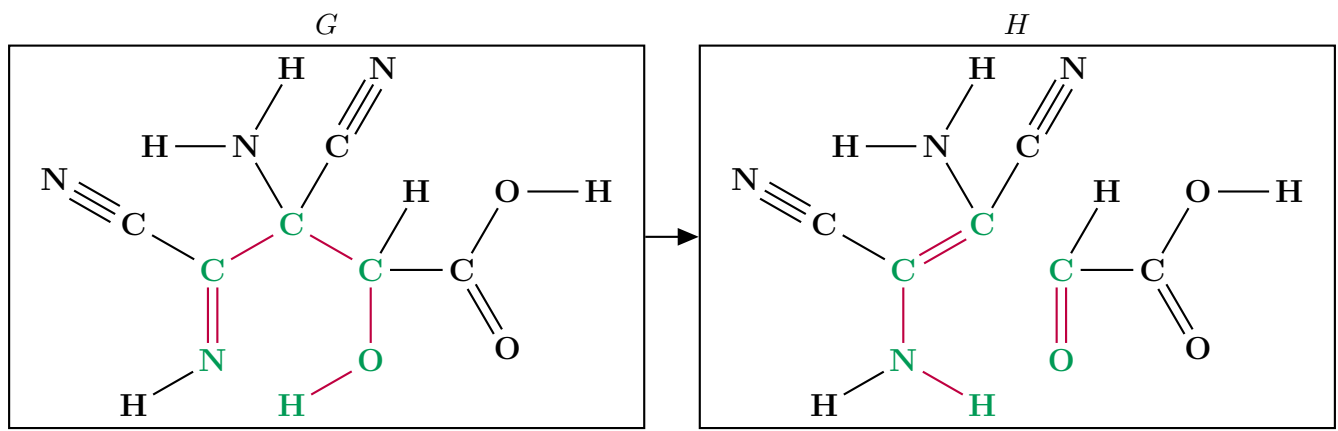
\begin{figure*}
  \centering
  \incFig{
  \begin{tikzpicture}[thick, node distance=20pt, vertex/.style={draw},
    normal/.style={->,>=triangle 45} ]
    \node[vertex,label=above:$G$](G) {\summaryRuleSide{figures/ruleG.pdf}};
    \node[vertex,label=above:$H$](H)[right=of G]
    {\summaryRuleSide{figures/ruleH.pdf}}; \draw[normal](G) to (H);
  \end{tikzpicture}
  }
  \caption[]{Graph Grammar derivation splitting the precursor $G$ in into a
    \chemfig{HCN}-tetramer and glyoxylate. The latter two molecules
    together form the graph $H$. The underlying production rule, a
    generalised reverse aldol addition \ruleAldolAdditionNInverse, is
    indicated by the coloured atoms and bonds so that $L$ is visible in $G$
    and $R$ is visible in $H$.}
  \label{fig:gg}
\end{figure*}

The application of a production to a graph $G$ is called a derivation. In a
compressed representation we can depict it as in Fig.~\ref{fig:gg}.  The
production is shown by the two highlighted subgraphs (red bonds and green
atoms) in the graphs denoted as $G$ and $H$. The highlighted patters in $G$
corresponds to an instance of precondition $L$, so that the production can
be applied. In $H$, thus, $L$ is replaced by $R$. The full DPO
representation of the rule application, along with the 18 chemical reaction
mechanisms considered in this contribution, can be found in the appendix
(the rule depicted in Fig.~\ref{fig:gg} is denoted as $r_{18}$ in the appendix).

An important technical consideration is that the graphs considered in our
framework are usually not connected. Chemical production rules indeed often
describe the formation of new bonds between two or even more educt
molecules or may lead to splitting of a single molecule into multiple
fragments. The graphs $G$ and $H$, hence, are not just molecules but rather
multi-sets of molecules taken from a certain chemical space. 

The application of a production of a given graph $G$ is a computationally
non-trivial task since it requires the solution of a subgraph isomorphism
problem namely that of finding $L$ as a subgraph of $G$. This is a
well-known NP-hard problem \cite{Cook:71}.  Nevertheless, it can
theoretically be solved efficiently for chemical graphs since they are
almost always planar and the pattern $L$ is small, see \citet{Eppstein:99}.
Our implementation uses the VF2 algorithm \cite{vf2} for this purpose.

\subsection{Generation and Exploration of Chemical Spaces}

The set of chemical reactions, modelled as production rules, can be applied
iteratively to a set of starting compounds. In principle, this procedure
generates the universe of all molecules and all reactions between them that
can be theoretically constructed from a given ``chemical repertoire''. In
each step, the chemical space grows. The production rules are intentionally
modelled in a generic way, such that the same rule can be applied to many
different combinations of compounds and even in many different ways on the
same set of compounds. The generative approach thus very quickly leads to
huge chemical spaces and the brute-force approach can only be applied for a
very limited number of steps. A practical issue in constructing a chemical
space is that we need to determine whether the result of a derivation is a
new molecule, and hence needs to be added to the space, or whether it is a
compound that we have already seen. This requires a large number of
solutions of the graph isomorphism problem.  For chemical systems, this
issue is usually dealt with by using canonical SMILES strings \cite{canSmiles}.
Since errors in the canonicalisation algorithm have
surfaced \cite{brokenSmiles}, we again use the VF2 algorithm \cite{vf2} to
verify results.

In order to control the computational effort, we employ methods from
programming language design in computer science and use these to define a
high-level strategy language. This will allow us to apply \emph{exploration
  strategies} and expand the chemical spaces in a much more controlled
way. For example, in contrast to a brute-force expansion, we can decide to
only apply a smaller subset of rules to particular subsets of compounds
specified in terms of certain graph-theoretical properties. Furthermore,
the framework allows us to prune uninteresting parts of the chemical space
by filtering the results of derivation again based on graph-theoretical
properties.  For a more formal introduction to the strategy framework we
refer to \citet{Andersen:14a}.

Eschenmoser \cite{esch07} presented several schemes to investigate the
relationship between HCN and the precursors of amino acids. To model some
of his schemes, we followed \emph{strict strategies}, i.e., we modelled
specific reactions paths based on an explicit sequence of rule
applications.  The reason for this restriction is that a brute-force
strategy in which all rules from the HCN chemistry are applied in an
iterative way expands the chemical space so quickly that the computational
resource are exhausted long before the products of interest are
reached. Still, even a strict strategy with a pre-determined order in which
production rules are applied, can lead to a plethora of different compounds
because each individual rule potentially can be applied in many different
ways. This follows from the fact that $L$ often admits several different
subgraph isomorphisms into $G$.

We also employ the very useful technique of an expansion step with a
subsequent \emph{closure} of the chemical space, i.e., we apply
(potentially all) chemical reactions to a set of chemical compounds, and
subsequently infer all possible reactions between all compounds. When
applying this strategy to a given network that fulfills a certain property
(e.g., being autocatalytic), then the space that results from an expansion
step with a subsequent closure operation can be used as an input to find
neighbouring solutions to the given specific chemical transformation
motif. How to find such motifs is discussed in the next section.

\subsection{Detection of Chemical Transformation Motives}

A chemical transformation motif is a subnetwork of a chemical space
with the following properties: (1) The motif has a well-defined
boundary to the outside, i.e., the educts, the products, and the food
molecules are predefined. (2) The motif is stoichiometrically
balanced, i.e., each chemical reactions contributes an integer flux
that reflects how often a specific reaction in the chemical space is
used within a specific pattern. Formally, the chemical spaces are
hypergraphs, and a chemical motif can be expressed as a hyperflow on
this hypergraph. Conceptually, integer hyperflows are similar to the
flows obtained from Flux Balance Analysis, see e.g.,
\citet{Orth:10}. The insistence on integer flows, however, allows to
infer mechanistically important properties. E.g., this approach allows
to distinguish between autocatalytic subsystems that need to be seeded by
their products and those that can ``auto-start''.

In order to bias the search for chemical transformation motifs or the
search for alternatives to predefined motifs, we use an objective
function. The probably most natural objective function will provide a maximum
parsimonious solution by minimizing the number of reactions that are used to realise
a sought transformation motif. Other possible objective functions include
the fluxes on the reactions, or they use a quantification for the reactions
that reflect the likelihood of reactions happening. For finding chemical
transformation motifs we formulate the question to be answered as an
Integer Linear Programming (ILP) problem. The technical details of this approach are
far beyond the scope of this paper and we refer to \citet{ourflowpaper}.

\section{Results}

In his conceptual paper \cite{esch07} Eschenmoser proposed to re-investigate
the non-robust parts of \chemfig{HCN} chemistry, and emphasises glyoxylate
as key a compound for the emergence of a primordial metabolism. A recent
experimental study explored the aldol-type chemistry of dihydroxyfumarate,
a tautomer of oxaloglycolate, and small aldehydes including glyoxylate \cite{Sagi:2012}.
 It demonstrates that under moderate-pH conditions (pH
7--8) in water, biologically relevant ketoses are formed selectively with
high yield. A follow-up study \cite{Butch:2013} revealed that, in contrast,
tartrate (CID~875) and oxalate (CID~971) are produced under high-pH conditions.
Dehydration and decarboxylation of tartrate, both well-known reactions,
readily yield oxaloacetate and pyruvate. Although high-pH conditions are
considered problematic in the prebiotic context, these studies nicely
illustrate that even under consistent basic conditions even the most
abundant products depend very sensitively on details of the reaction
conditions. \textit{In silico} explorations of the chemical space in the
vicinity of experimentally known or theoretically inferred reaction
sequences may drastically speed up the search for novel and unconventional
routes to biologically relevant molecules. To demonstrate this point, we
modelled the necessary \chemfig{HCN} reaction chemistry proposed in
Eschenmoser's paper \cite{esch07} with the help of our graph grammar
framework. In this manner we can explicitly construct the reaction network
surrounding the suggested pathways, so that we can explore potential
alternatives in the chemical space where these routes are embedded. The
hypothetical reaction sequences generated by the graph grammar framework can
be refined whenever experimental evidence is collected. The 18 reaction
types used in this contribution are listed in the appendix. We have
deliberately specified as little context as possible to guarantee a broad
applicability.

\subsection{Overview}

\begin{figure*}
	\centering
	\incFig{
\newcommand{\molFig}[1]{\includegraphics[scale=0.35]{namedMols/#1}}
\newcommand{\CataReaction}[9]{
	\draw[pathMiddle, #9] (#2) to[#5] (#1);
	\draw[path, #9] (#1) to[#6] (#2);
	\draw[pathMiddle, #9] (#3) to[#7] (#1);
	\draw[path, #9] (#1) to[#8] (#4);
}
{\scriptsize
\begin{tikzpicture}[
	hnode/.style={},
	reaction/.style={draw},
	path/.style={->,>=stealth', dashed},
	pathMiddle/.style={dashed}
]
\node[hnode] (pyr) {\molFig{pyr}};
\node[hnode, right=3em of pyr] (oaa) {\molFig{oaa}};
\node[hnode, right=3em of oaa] (glycolaldehyde) {\molFig{glycolaldehyde}};
\node[anchor=south] (labelTop) at (oaa.90) {Oxaloacetate};
\path (labelTop) -|
		node[text depth=0] {Glycolaldehyde}
	(glycolaldehyde);
\path (labelTop) -|
		node[text depth=0] {Pyruvate}
	(pyr);

\node[hnode, below=of glycolaldehyde, label=165:{Oxaloglycolate}] (dihydroxymalate) {\molFig{dihydroxymalate}};
\node[hnode, below=of dihydroxymalate] (oxoaspartate) {\molFig{oxoaspartate}};
\node[hnode, below=of oxoaspartate] (damnHydroHydro) {\molFig{damnHydroHydro}};
\node[hnode, left=5em of damnHydroHydro] (damnHydro) {\molFig{damnHydro}};
\node[hnode, left=5em of damnHydro, label=left:{DAMN}] (damn) {\molFig{damn}};
\node[hnode, above=of damn] (trimer) {\molFig{trimer}};
\node[hnode, above=of trimer] (dimer) {\molFig{dimer}};
\node[hnode, above=of dimer] (hcn) {\molFig{hcn}};

\path (hcn) to
		node[hnode, xshift=-10em] (glxDamn) {\molFig{glx}}
		node[reaction, xshift=-5em] (hcnDamnCata) {}
	(damn);
\node[anchor=south] at (glxDamn.90) {Glyoxylate};
\CataReaction{hcnDamnCata}{glxDamn}{hcn}{damn}
	{out=45, in=90}
	{in=-45, out=-90}
	{out=-145, in=80}
	{in=130, out=-80}{}

\node[hnode, above=10em of damnHydro, label=87:{Glyoxylate}] (glx) {\molFig{glx}};
\draw[path] (dimer) to[bend right=10] (glx);
\path (dimer) to[bend left=10]
		node[reaction, glxAutocataStyle] (dimerGlxCata) {}
	(glx);
\CataReaction{dimerGlxCata}{glx}{dimer}{glx}
	{out=130, in=100}
	{bend right=5}
	{bend left=5}
	{bend left=5}{glxAutocataStyle}

\path (damn) to[bend left=40]
		node[reaction] (damnDamnHydroCata) {}
	(damnHydro);
\CataReaction{damnDamnHydroCata}{glx}{damn}{damnHydro}
	{bend right=10}
	{bend right=10}
	{bend left=10}
	{bend left=10}{}
	
\path (damnHydro) to[bend left=40]
		node[reaction] (damnHydroDamnHydroHydroCata) {}
	(damnHydroHydro);
\CataReaction{damnHydroDamnHydroHydroCata}{glx}{damnHydro}{damnHydroHydro}
	{bend right=10}
	{bend right=10}
	{bend left=10}
	{bend left=10}{}

\path (oxoaspartate) to[bend left=80]
		node[reaction, dihydroxymalateAutocataStyle] (oxoaspartateDihydroxymalateCata) {}
	(dihydroxymalate);
\CataReaction{oxoaspartateDihydroxymalateCata}{dihydroxymalate}{glx}{oxoaspartate}
	{bend right=10}
	{bend right=10}
	{bend left=5}
	{bend left=5}{dihydroxymalateAutocataStyle}

\draw[edge] (hcn) to (dimer);
\draw[edge] (dimer) to (trimer);
\draw[path] (trimer) to (damn);
\draw[edge] (damn) to (damnHydro);
\draw[edge] (damnHydro) to (damnHydroHydro);
\draw[path] (damnHydroHydro) to (oxoaspartate);
\draw[path, dihydroxymalateAutocataStyle] (oxoaspartate) to (dihydroxymalate);
\draw[path] (dihydroxymalate) to (oaa);
\draw[path] (dihydroxymalate) to (glycolaldehyde);
\draw[path] (oaa) to (pyr);
\end{tikzpicture}
}
	}
	\caption{Overview of Eschenmoser's hypothetical chemical space.}
	\label{fig:overview}
\end{figure*}
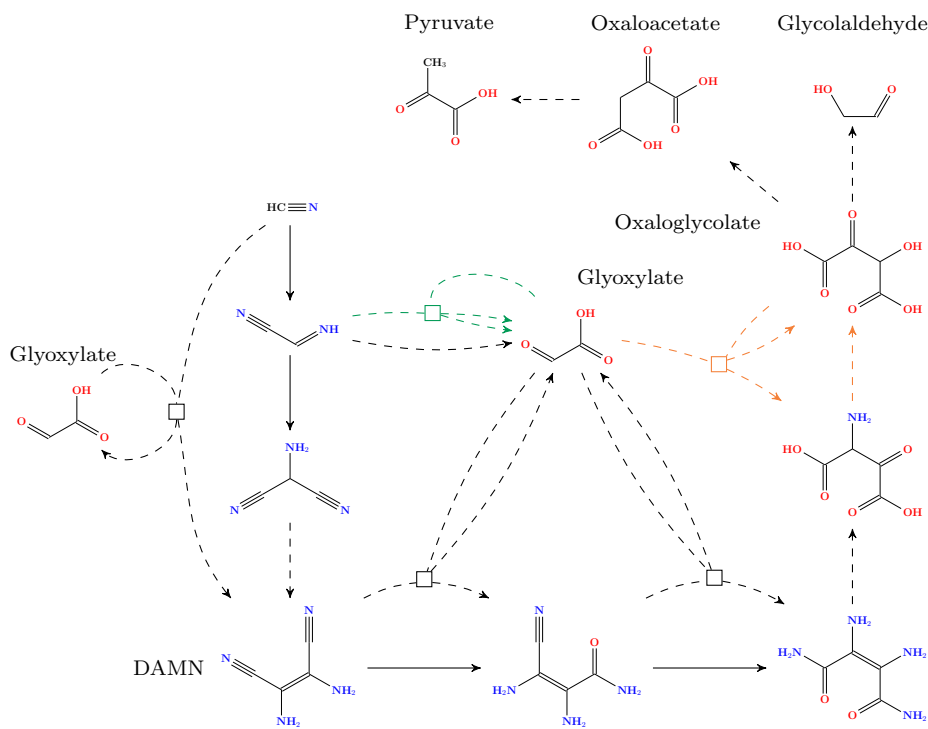

Fig.~\ref{fig:overview} summarises our contribution. It merges several of
Eschenmoser's schemes \cite{esch07} and integrates them with our
findings. The figure shows only key compounds and is intended to facilitate
the navigation through the much more detailed analysis of the individual
components that we have modelled and analysed with our graph
grammar framework.

All results given in this section are based on the automatic computational
inference of chemical transformation motifs within a large chemical
space. Solutions of the various underlying optimisation questions
correspond to automatically inferred detailed descriptions of the
corresponding mechanism for the question asked. These computational results
contain a plethora of additional information that we neither use nor
discuss in this paper. For example, a complete set of atom maps is obtained
as a by-product. From these, one could immediately infer atom traces for
specific isotope labelling experiments to distinguish alternative pathways
\cite{Andersen:14b}.

In Fig.~\ref{fig:overview}, any edge drawn as a solid line indicates that a
single production rule, i.e., a single reaction has been used. Dashed edges
or hyperedges denote larger chemical subspaces, i.e., sub-hypergraphs. The
two dashed orange (hyper-)edges refer to the hypothetical autocatalytic
subnetwork of oxaloglycolate, using glyoxylate as food molecule. Such a
hypothetical autocatalytic cycle has been presented and discussed in detail
by Eschenmoser \cite{esch07}. Below, we will discuss alternative routes
found by computational inference. The dashed green hyperedge depicts the
hypothetical autocatalytic subnetwork of Glyoxylate, using \chemfig{HCN} as
food molecule. Such a subsystem has not been proposed in earlier work. In
section \ref{sect:autoglx} we will discuss several alternative co-optimal
solutions for this chemical transformation motif.

Oxaloglycolate and its oligomeres are precursors of carbohydrates and
$\alpha$-keto acids, which themselves are potential starting compounds for a
primordial metabolism. In Fig.~\ref{fig:overview} oxaloglycolate is
depicted as a precursor to glycolaldehyde, the dashed line corresponds to
four subsequent production rules that correspond to a decarboxylation. 
Furthermore, via two production rule steps oxaloglycolate is
reduced to oxaloacetate, which in two subsequent steps leads to pyruvate.

\subsection{Pathways from the HCN-tetramer to Oxaloglycolate}

\begin{figure*}
  \centering
  \incFig{
  \newcommand{\modIncludeDir}{productLikeGraph}
  \resizebox{\textwidth}{!}{ \input{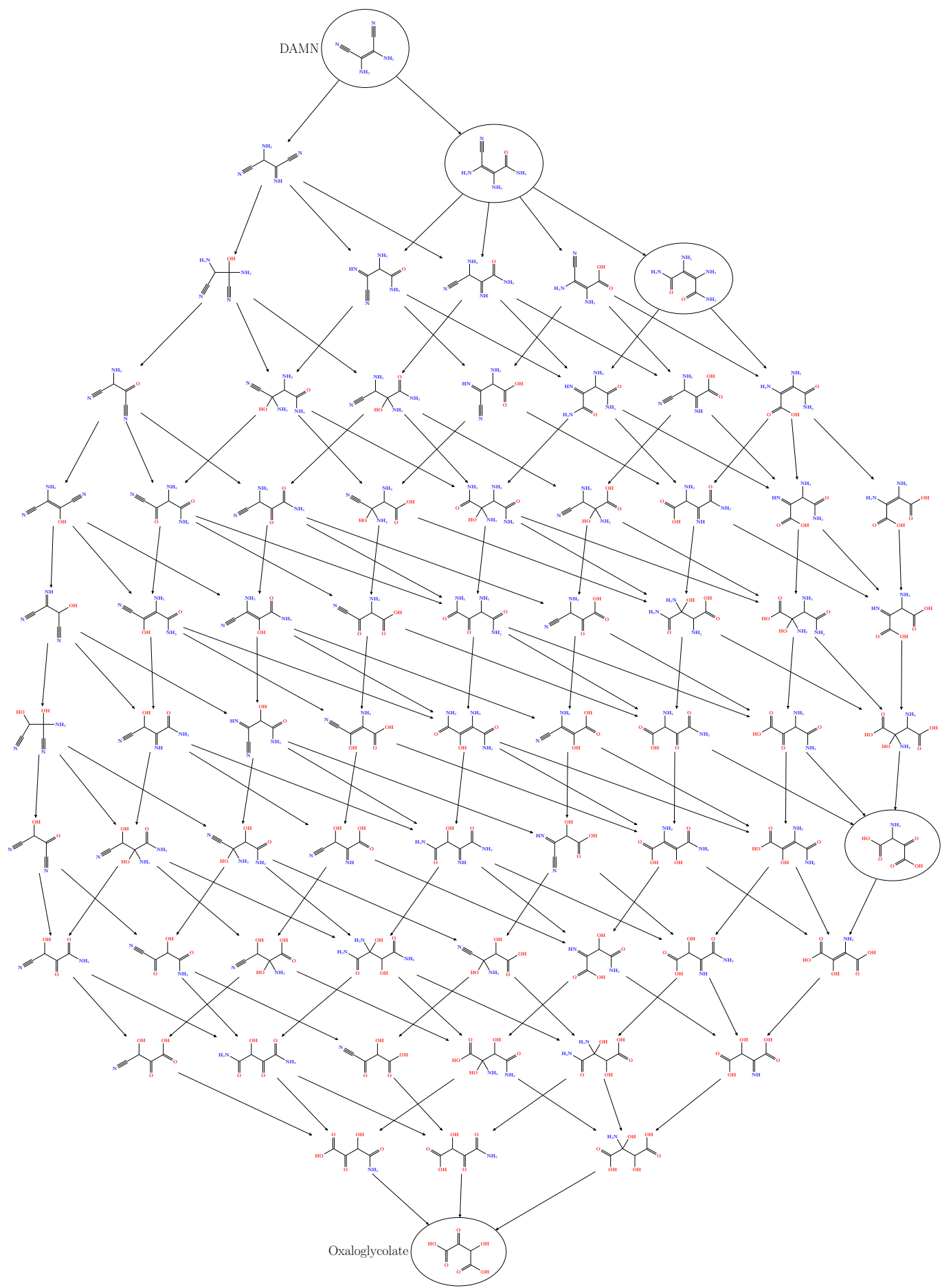} }
  }
  \caption{A superposition of optimal pathways from the HCN-tetramer to
    oxaloglycolate, illustrating the combinatorial complexity for even small
    subspaces.}
  \label{fig:product}
\end{figure*}

At first glance the conversion of the \chemfig{HCN}-tetramer into
oxaloglycolate seems to be a straightforward hydrolysation
process. Eschenmoser's paper \cite{esch07} and Fig.~\ref{fig:overview}
illustrates this process in strongly abstracted form with just 3 steps. A
mechanistic model, however, shows that 11 steps are
required. Fig.~\ref{fig:product} summarises a superposition of all pathways
of minimum length starting at DAMN and ending in oxaloglycolate. The figure
emphasises the combinatorial nature of the \chemfig{HCN} hydrolysation
chemistry.  Although all connections in the graph in
Fig.~\ref{fig:overview} seem to be 1-to-1, this is in fact not the
case. The simple molecules \chemfig{H_2O}, \chemfig{NH_3}, and
\chemfig{HCN} were suppressed in the drawing to reduce cluttering. The
intermediate compounds selected as representatives in the overview figure
(see Fig.~\ref{fig:overview}) are highlighted in Fig.~\ref{fig:product}. A closer
inspection of the structure of the network in Fig.~\ref{fig:product} shows
that it closely resembles a Cartesian graph product\cite{sa60}. This feature is 
based on the fact that the temporal order of several of the
intermediate steps can be permuted. A consequence of the product-like
structure is a high level of confluence, i.e., the fact that a large number
of partially overlapping alternative pathways lead from the same educts to
the same products. The hydrolysation chemistry of DAMN behaves like a
dynamic combinatorial library that can adjust the internal fluxes upon
change of the source or sink reactions.

\subsection{The Autocatalytic Production of Glyoxylate}
\label{sect:autoglx}

Eschenmoser \cite{esch07} discussed the usage of glyoxylate as a starting
material for the autocatalytic production of oxaloglycolate. Glyoxylate
itself is produced via slow hydrolysation chemistry from the
\chemfig{HCN}-dimer imino\-aceto\-nitrile (CID 14496737). The production of oxaloglycolate
and a potential downstream primordial metabolism therefore also depends on
\chemfig{HCN} as food source. If an alternative pathway for the production
of glyoxylate becomes accessible, the dependent downstream metabolism
remains unaffected and formally adapts to the new precursor of glyoxylate
as food source. However, the production of glyoxylate from its precursor
molecule is the rate-limiting step for the downstream processes. An
autocatalytic cycle feeding on a slow hydrolysation process can not
exhibit exponential growth characteristics and thus cannot canalise the
reaction network.  Eschenmoser therefore asked how glyoxylate could be
produced more efficiently from its precursor compound(s).

\begin{figure*}
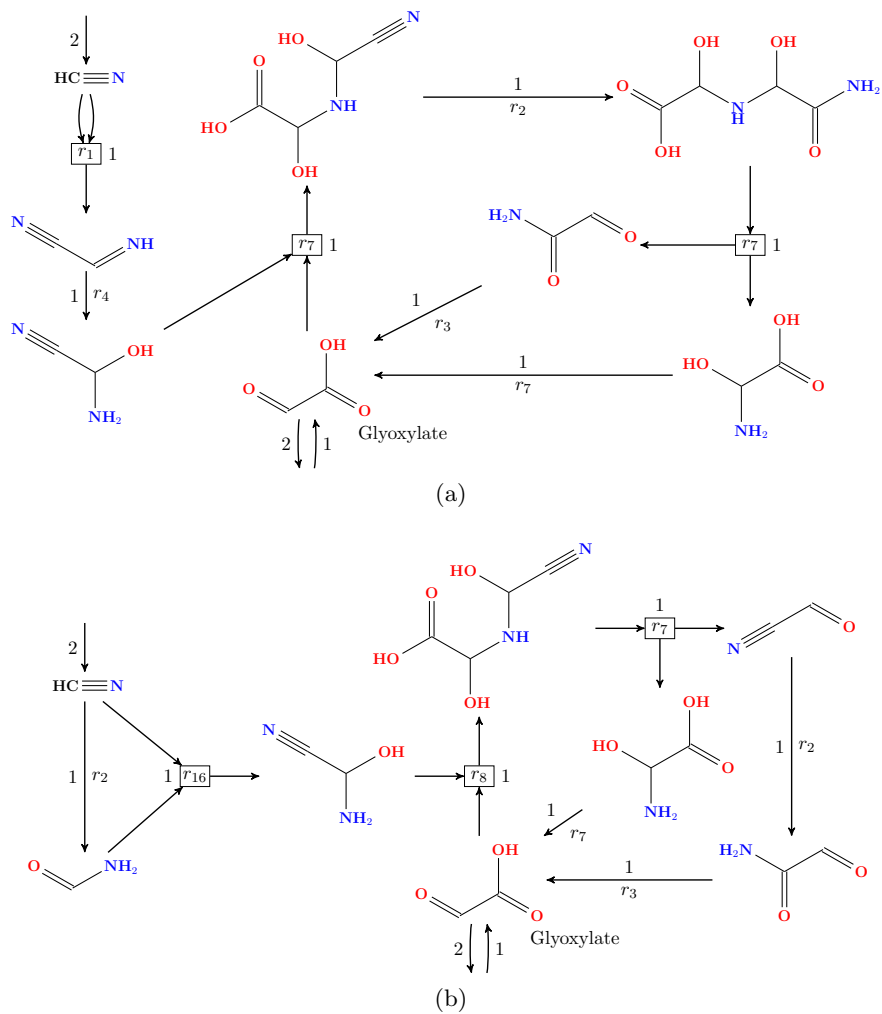

  \centering
  \newcommand{\modIncludeDir}{glxAutocata}
  \subfloat[]{
  	\incFig {
     \scalebox{0.65}{
       \input{\modIncludeDir/010_dg_0_1111_f_0_32_filt.tex}
     }
     }
     \label{fig:glx1}
   }
  
  \subfloat[]{
  	\incFig{
    \scalebox{0.65}{
      \input{\modIncludeDir/018_dg_0_1111_f_0_6_filt.tex}
    }
    	}
     \label{fig:glx2}
  }
  \caption{Two autocatalytic pathways for the autocatalytic
    production of glyoxylate. Hyperedges are labelled with the
    reaction id and the integer flux value of the specific reaction. }
  \label{fig:glxautocata}
\end{figure*}

We searched for potential autocatalytic cycles that produce
glyoxylate from \chemfig{HCN}. Even in the very strict modelling approach we
applied, several autocatalytic solutions were found. Two arbitrarily
chosen solutions with a minimum number of reactions have been analysed in
more detail, see Fig.~\ref{fig:glxautocata}. The \chemfig{HCN}
hydrolysation chemistry contains many more solutions in particular if the
constraint on the minimality of the autocatalytic cycle is relaxed. Both
solutions produce two glyoxylate molecules from a single glyoxylate (acting
as autocatalyst) and \chemfig{HCN} as food source. The solutions are
depicted as hyperflows, i.e., any reaction has an assigned integer value
(shown as edge label). As usual \chemfig{H_2O} and \chemfig{NH_3} (with a
balance of $+3$ and $-2$, resp.), are suppressed from the drawing. The food
molecule for both autocatalytic cycles are produced via slightly different
pathways (Fig.~\ref{fig:glx1} via iminoacetonitrile, Fig.~\ref{fig:glx2} via formamide). Although the chemistry
in the two cycles differ, the last step which produces two copies of
glyoxylate are identical in both solutions. It involves amide hydrolysis
which, depending on the reaction conditions, can be a slow process. During
one turnover, both cycles consume two \chemfig{HCN} molecules. We want to
emphasise that these solutions only possess the proper topology for being
autocatalytic cycles. We do not make any assumption about the kinetic
rates or potential draining reactions which both can dampen the flux around
the cycle to such an extend, that the autocatalytic behaviour is
lost. However, our approach would allow to include such properties via
additional constraints or an adequate modification of the objective
function that is used in our optimisation approach.

\subsection{The Autocatalytic Production of Oxaloglycolate}

\begin{figure*}
  \centering
  \incFig{
  \newcommand{\modIncludeDir}{doubleAutocata}
  \resizebox{\textwidth}{!}{
    \input{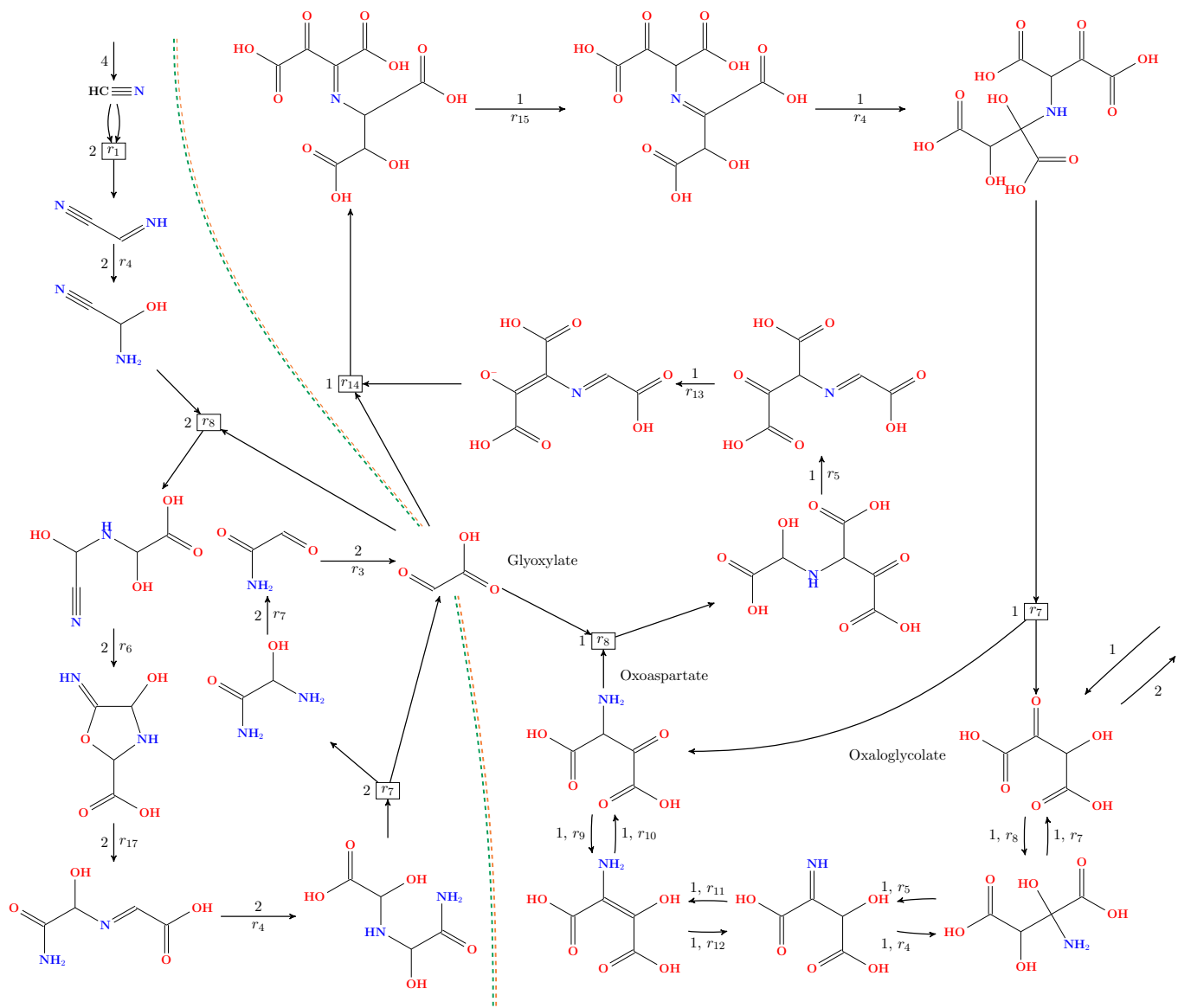}
  }
  }
  \caption{The autocatalytic production of oxaloglycolate, using glyoxylate as a
    food source (to the right of the separating green-orange dashed line). This chemical
    space follows Eschenmoser's suggestions and integrates a subspace for
    the autocatalytic production of the glyoxylate (left of the
    green-orange dashed line) as found in Section \ref{sect:autoglx}.}
      \label{fig:combined}
\end{figure*}

Eschenmoser \cite{esch07} proposed an hypothetical autocatalytic cycle, that
produces oxaloglycolate from glyoxylate. This transformation requires that
the glyoxylate dimer or its amides acts as ``umpolung-catalyst''. With the
help of the strict expansion strategy we reproduced this model
computationally. The corresponding autocatalytic cycle is depicted in the
overview figure (simplified in Fig.~\ref{fig:overview}, as two dashed
orange (hyper-)edges). The right-hand side of Fig.~\ref{fig:combined} shows the
detailed reaction pathway generated by the graph grammar approach. More
precisely, it represents an optimal hyperflow in the chemical space in
which both glyoxylate and oxaloglycolate are produced autocatalytically.
This pathway illustrates the coupling of two autocatalytic cycles, one
producing oxaloglycolate from glyoxylate and the other producing glyoxylate
from \chemfig{HCN}. The overall pathway consumes 4 \chemfig{HCN} in order
to produce one oxaloglycolate.

A cascading of autocatalytic reaction cycles has the advantage, that the
downstream cycles are not rate limited by the upstream processes that
produce their food molecules, and that symmetry-breaking of a homogeneous
mixture can be very fast. In our particular case, when the autocatalytic
cycle of the glyoxylate production kicks in, all the molecular mass
distributed over the large reversible \chemfig{HCN}-hydrolysation network
(Fig.~\ref{fig:product}) is re-absorbed into the autocatalytic glyoxylate
cycle. The effect of this ``flux focusing'' is an efficient re-routing of
potentially lost raw material towards the downstream primordial metabolic
processes.

\begin{figure*}
\centering
\incFig{
\newcommand{\modIncludeDir}{doubleAutocataShorter}
\resizebox{\textwidth}{!}{
	\input{\modIncludeDir/025_dg_0_1111_f_2_0_filt.tex}
}
}
\caption[]{
  One of the optimal subspaces within which oxaloglycolate is produced
  autocatalytically from \chemfig{HCN} as food source and the in which the
  intermediate food source glyoxylate is itself is produced in an
  autocatalytic manner. To find this solution the space depicted in
  Fig.~\ref{fig:combined} was expanded and a closure operation was applied
  to the expanded space.  The two autocatalytic cycles are separated by the
  green-orange dotted line, cmp. Fig.~\ref{fig:combined}.}
  \label{fig:combined-esch}
\end{figure*}

At this point the question arises if there are alternative cascades of
autocatalytic cycles hidden in the chemical space using glyoxylate and
oxaloglycolate? In order to search for such alternatives we expanded the
chemical space spanned by the hypergraph underlying Fig.~\ref{fig:combined}
and then applied a closure operation, which embeds the reaction network
employed for Fig.~\ref{fig:combined} into its nearby ``shadow
network''. The resulting chemical space is composed of 415 compounds
connected by 1116 reactions (compared to the network of 95 compounds and
133 reactions, which was used as the underlying network for inferring the
chemical transformation motif depicted in Fig.~\ref{fig:combined}).

The objective function used for the inference of Fig.~\ref{fig:combined}
was applied to this larger chemical space to search for alternative
autocatalytic cascades. A solution is depicted in
Fig.~\ref{fig:combined-esch}. The number of reactions in the alternative
solution dropped from 24 to 22. Notice that both
autocatalytic sub-networks are slightly different from the solution
depicted in Fig.~\ref{fig:combined}. The splitting step of the glyoxylate
cycle, which generates the two copies of glyoxylate uses only the
break-down of a hemiaminal, which can be considered to be a fast
reaction. In the oxaloglycolate cycle the addition step of the first
glyoxylate goes via a different tautomer of oxoaspartate. Although the
oxaloglycolate sub-network was combinatorially modified, the number of
reactions stayed the same. However, even one expansion step followed by a
closure operation reduced the number of reactions used in the glyoxylate
sub-network from 9 to 7.

\section{Discussion and Conclusions}

The analysis of autocatalytic subsystems of chemical reaction networks has
become a lively research area in prebiotic chemistry \cite{pe12}.
Autocatalysis has the potential to canalise reactions in such a way that a
relatively small number of well-defined organic molecules can accumulate in
substantial concentrations. The key interest is, of course, in the
production of the current constituents of living cells and in pathways that
may feed core subsystems of present-day metabolisms such as the reductive
citric acid cycle \cite{Smith:04}. A good example of such
hypothetical autocatalytic feeding networks are those that start from the
``formal hydrates of \chemfig{CO}'', in particular glyoxylate and
2,3-hydroxyfumarate.

The scenario of prebiotic synthesis proposed by Albert Eschenmoser
\cite{esch07} is one of the most popular ones of this type. Other
hypotheses worth mentioning in this context are, e.g., a model for the
cyclic production of \chemfig{HCN} \cite{schwartz82} and a model of sugar
chemistry \cite{weber07}. A key feature of Eschenmoser's model is the idea
that the main ``food sources'', and hence the main pathways, have evolved
in a step-wise manner. Transitions between dominating reaction pathways are
triggered whenever intermediate compounds become available that
``kick-start'' new autocatalytic cycles leading to their own
production. In Eschenmoser's model, first \chemfig{HCN} is used as a food
source in order to produce glyoxylate. Later-on glyoxylate is consumed as a
food source for the autocatalytic production of oxaloglycolate, using
itself as an (umpolung-)catalyst.

In order to study the possible chemical pathways in a more systematic
manner efficient computational approaches are required to deal with the
combinatorial nature and the immense size of chemical spaces. This in turn
calls for a modelling approach that has a solid formal background. Here we
use a graph-theoretical approach, more precisely graph transformation
systems. These make it possible to generate well-defined subsets of
chemical spaces. This type of computational modelling indeed allows the
exploration of large numbers of alternative pathways, some of which are
rather unexpected. In a recent investigation into
\chemfig{HCN}-polymerisation/-hydrolysis chemistry \cite{Andersen:13b}, for
example, we found several plausible alternatives to Or{\'o}'s prebiotic
adenine synthesis \cite{Oro:61}. In the computational re-analysis of
Eschenmoser's hypothesis outlined here, we find that autocatalytic pathways
can be found abundantly, often involving the same key intermediates. 

The formal approach not only recovers the proposals put forward by
experienced chemists but also reveals a plethora of alternatives that
nevertheless match the same chemical transformation motives as those
proposed in Eschenmoser's work. 

In some cases, such as the production of glyoxylate from \chemfig{HCN} we
found previously undescribed autocatalytic production cycles (see
Fig.~\ref{fig:glxautocata} and Fig.~\ref{fig:combined}).  In other parts of
the network we found plausible alternative pathways such as the
autocatalytic umpolung cycle for the oxaloglycolate production (see
Fig.~\ref{fig:combined-esch}). Some of these solutions are particularly
appealing since they involve even fewer reaction steps than their
previously described alternatives. The superposition of the optimal
pathways from the \chemfig{HCN}-tetramer to oxaloglycolate,
Fig.~\ref{fig:product}, highlights a product like organisation of a large
number of confluent reaction sequences. The analysis of the shadow, i.e.,
the immediate vicinity of the union of the networks autocatalytic in
glyoxylate and oxaloglycolate, illuminates a complex interplay of alternative
routes.

Of course, the combinatorial analysis of the network only describes
possibilities and plausibilities. Considerations of thermodynamic stability
and in particular of the kinetics of alternative reactions have to be
invoked to further narrow down the scenarios to those that are good
candidates for having taken place on Early Earth. Nevertheless, the
generation of the chemical spaces remains a necessary first step for any
form of more refined and more realistic model of prebiotic chemistry. 

\section*{Acknowledgments}

This work was supported in part by the EU-FET grant RiboNets 323987,
the COST Action CM1304 ``Emergence and Evolution of Complex Chemical
Systems'', and the Danish Council for Independent Research, Natural Sciences.

\fancyhf{}
\rhead{ \fancyplain{}{REFERENCES}}
\bibliographystyle{abbrvnat}
\bibliography{eschenmoser}

\newpage
\fancyhf{}
\rhead{ \fancyplain{}{APPENDIX}}
\section*{Appendix}
\subsection*{Graph Grammar Rules}
The following list contains $18$ chemical reactions modelled as graph
grammar production rules. A production rule in the DPO framework is
usually denoted as $r = (L\xleftarrow{} K\xrightarrow{} R)$, where
each rule $r$ consist of three graphs (fragments of molecules) $L$,
$R$, and $K$ called the left, right, and context graph,
respectively. Informally, the left graph $L$ can be replaced by a
right graph $R$ if the corresponding subgraphs were found (see section
\ref{sect:gg}). The indices of the production rules are consistent to
the indices of the production rules as used throughout the paper.

{
\removeStuff{
\renewcommand{\summaryRuleSide}[1]{
        \begin{minipage}{0.28\textwidth}
                \begin{center}
                        \fitpic{#1}
                \end{center}
        \end{minipage}
}
}

\newcommand{\summaryRuleSpan}[2]{
    \begin{center}
    \incFig{
    \begin{tikzpicture}[thick,
        node distance=20pt,
        vertex/.style={draw},
        normal/.style={->,>=triangle 45} 
    ]   
    \node[vertex,label=above:$L$](L)                {\summaryRuleSide{#1.L}};
    \node[vertex,label=above:$K$](K)[right=of L]    {\summaryRuleSide{#1.K}};
    \node[vertex,label=above:$R$](R)[right=of K]    {\summaryRuleSide{#1.R}};
    \draw[normal](K) to (L);
    \draw[normal](K) to (R);
    \end{tikzpicture}
    }
    \end{center}
}

\newcommand{\summaryRule}[8]{%
    \subsubsection*{#1}
    \summaryRuleSpan{#4}{#5}
}

\newcommand{\modIncludeDir}{rules}
\summaryRule{\ruleAddHCN, Add HCN}{\modIncludeDir/002_r_7.1010000}{\modIncludeDir/002\_r\_7.1010000}{\modIncludeDir/003_r_7.1110010}{\modIncludeDir/003\_r\_7.1110010}{\modIncludeDir/004_r_7_combined}{\modIncludeDir/004\_r\_7\_combined}{\modIncludeDir/005_r_7_constraints.tex}
\summaryRule{\ruleCNToAmid, CN to Amide}{\modIncludeDir/007_r_11.1010000}{\modIncludeDir/007\_r\_11.1010000}{\modIncludeDir/008_r_11.1110010}{\modIncludeDir/008\_r\_11.1110010}{\modIncludeDir/009_r_11_combined}{\modIncludeDir/009\_r\_11\_combined}{\modIncludeDir/010_r_11_constraints.tex}
\summaryRule{\ruleAmidToAcidWater, Amide to Acid}{\modIncludeDir/012_r_12.1010000}{\modIncludeDir/012\_r\_12.1010000}{\modIncludeDir/013_r_12.1110010}{\modIncludeDir/013\_r\_12.1110010}{\modIncludeDir/014_r_12_combined}{\modIncludeDir/014\_r\_12\_combined}{\modIncludeDir/015_r_12_constraints.tex}
\summaryRule{\ruleWaterToImine, Water to Imine}{\modIncludeDir/022_r_17.1010000}{\modIncludeDir/022\_r\_17.1010000}{\modIncludeDir/023_r_17.1110010}{\modIncludeDir/023\_r\_17.1110010}{\modIncludeDir/024_r_17_combined}{\modIncludeDir/024\_r\_17\_combined}{\modIncludeDir/025_r_17_constraints.tex}
\summaryRule{\ruleWaterToImineInverse, Imine to Water}{\modIncludeDir/027_r_18.1010000}{\modIncludeDir/027\_r\_18.1010000}{\modIncludeDir/028_r_18.1110010}{\modIncludeDir/028\_r\_18.1110010}{\modIncludeDir/029_r_18_combined}{\modIncludeDir/029\_r\_18\_combined}{\modIncludeDir/030_r_18_constraints.tex}
\summaryRule{\ruleAddAlcohol, Add Alcohol}{\modIncludeDir/037_r_25.1010000}{\modIncludeDir/037\_r\_25.1010000}{\modIncludeDir/038_r_25.1110010}{\modIncludeDir/038\_r\_25.1110010}{\modIncludeDir/039_r_25_combined}{\modIncludeDir/039\_r\_25\_combined}{\modIncludeDir/040_r_25_constraints.tex}
\summaryRule{\ruleAminal, Aminal}{\modIncludeDir/042_r_29.1010000}{\modIncludeDir/042\_r\_29.1010000}{\modIncludeDir/043_r_29.1110010}{\modIncludeDir/043\_r\_29.1110010}{\modIncludeDir/044_r_29_combined}{\modIncludeDir/044\_r\_29\_combined}{\modIncludeDir/045_r_29_constraints.tex}
\summaryRule{\ruleAminalInverse, Aminal (Inverse)}{\modIncludeDir/047_r_30.1010000}{\modIncludeDir/047\_r\_30.1010000}{\modIncludeDir/048_r_30.1110010}{\modIncludeDir/048\_r\_30.1110010}{\modIncludeDir/049_r_30_combined}{\modIncludeDir/049\_r\_30\_combined}{\modIncludeDir/050_r_30_constraints.tex}
\summaryRule{\ruleKetoEnol, Keto to Enol}{\modIncludeDir/062_r_34.1010000}{\modIncludeDir/062\_r\_34.1010000}{\modIncludeDir/063_r_34.1110010}{\modIncludeDir/063\_r\_34.1110010}{\modIncludeDir/064_r_34_combined}{\modIncludeDir/064\_r\_34\_combined}{\modIncludeDir/065_r_34_constraints.tex}
\summaryRule{\ruleEnolKeto, Enol to Keto}{\modIncludeDir/067_r_35.1010000}{\modIncludeDir/067\_r\_35.1010000}{\modIncludeDir/068_r_35.1110010}{\modIncludeDir/068\_r\_35.1110010}{\modIncludeDir/069_r_35_combined}{\modIncludeDir/069\_r\_35\_combined}{\modIncludeDir/070_r_35_constraints.tex}
\summaryRule{\ruleKetimineEnolimine, Ketimine to Enolimine}{\modIncludeDir/077_r_40.1010000}{\modIncludeDir/077\_r\_40.1010000}{\modIncludeDir/078_r_40.1110010}{\modIncludeDir/078\_r\_40.1110010}{\modIncludeDir/079_r_40_combined}{\modIncludeDir/079\_r\_40\_combined}{\modIncludeDir/080_r_40_constraints.tex}
\summaryRule{\ruleEnolimineKetimine, Enolimine to Ketimine}{\modIncludeDir/082_r_41.1010000}{\modIncludeDir/082\_r\_41.1010000}{\modIncludeDir/083_r_41.1110010}{\modIncludeDir/083\_r\_41.1110010}{\modIncludeDir/084_r_41_combined}{\modIncludeDir/084\_r\_41\_combined}{\modIncludeDir/085_r_41_constraints.tex}
\summaryRule{\ruleKetoEnolUmpolung, Keto to Enol (Umpolung)}{\modIncludeDir/087_r_42.1010000}{\modIncludeDir/087\_r\_42.1010000}{\modIncludeDir/088_r_42.1110010}{\modIncludeDir/088\_r\_42.1110010}{\modIncludeDir/089_r_42_combined}{\modIncludeDir/089\_r\_42\_combined}{\modIncludeDir/090_r_42_constraints.tex}
\summaryRule{\ruleAdditionUmpolung, Addition (Umpolung)}{\modIncludeDir/092_r_43.1010000}{\modIncludeDir/092\_r\_43.1010000}{\modIncludeDir/093_r_43.1110010}{\modIncludeDir/093\_r\_43.1110010}{\modIncludeDir/094_r_43_combined}{\modIncludeDir/094\_r\_43\_combined}{\modIncludeDir/095_r_43_constraints.tex}
\summaryRule{\ruleCNCEnolSwap, CNC Enol Swap}{\modIncludeDir/097_r_44.1010000}{\modIncludeDir/097\_r\_44.1010000}{\modIncludeDir/098_r_44.1110010}{\modIncludeDir/098\_r\_44.1110010}{\modIncludeDir/099_r_44_combined}{\modIncludeDir/099\_r\_44\_combined}{\modIncludeDir/100_r_44_constraints.tex}
\summaryRule{\ruleAddHCNToAldehyde, Add HCN to Aldehyde}{\modIncludeDir/102_r_45.1010000}{\modIncludeDir/102\_r\_45.1010000}{\modIncludeDir/103_r_45.1110010}{\modIncludeDir/103\_r\_45.1110010}{\modIncludeDir/104_r_45_combined}{\modIncludeDir/104\_r\_45\_combined}{\modIncludeDir/105_r_45_constraints.tex}
\summaryRule{\ruleEtherBreak, Ether Break}{\modIncludeDir/107_r_47.1010000}{\modIncludeDir/107\_r\_47.1010000}{\modIncludeDir/108_r_47.1110010}{\modIncludeDir/108\_r\_47.1110010}{\modIncludeDir/109_r_47_combined}{\modIncludeDir/109\_r\_47\_combined}{\modIncludeDir/110_r_47_constraints.tex}
\summaryRule{\ruleAldolAdditionNInverse, Aldol Addition N (Inverse)}{\modIncludeDir/072_r_39.1010000}{\modIncludeDir/072\_r\_39.1010000}{\modIncludeDir/073_r_39.1110010}{\modIncludeDir/073\_r\_39.1110010}{\modIncludeDir/074_r_39_combined}{\modIncludeDir/074\_r\_39\_combined}{\modIncludeDir/075_r_39_constraints.tex}
}

\subsection*{Double Pushout Diagram for Figure~\ref{fig:gg}}
This section givens an example of using production rule
\ruleAldolAdditionNInverse\ for the derivation from an educt (graph
$G$) to the two products (graph $H$, which represents two chemical
compounds, namely the HCN-tetramer and glyoxylate).  The edges changed
by the transformation are shown in red and the vertices from $K$ are
shown in green. All arrows represent embeddings of one graph into
another, formally a subgraph monomorphism.

{
\newcommand{\modIncludeDir}{dpo}
\newcommand{\summaryDerivation}[4]{
    \begin{center}
    \incFig{
    \begin{tikzpicture}[thick,
        node distance=20pt,
        vertex/.style={draw},
        normal/.style={->,>=triangle 45} 
    ]   
    \node[vertex,label=above:$L$](L)                {\summaryRuleSide{#1.derL}};
    \node[vertex,label=above:$K$](K)[right=of L]    {\summaryRuleSide{#1.derK}};
    \node[vertex,label=above:$R$](R)[right=of K]    {\summaryRuleSide{#1.derR}};
    \draw[normal](K) to (L);
    \draw[normal](K) to (R);
    \node[vertex,label=below:$G$](G)[below=of L]    {\summaryRuleSide{#3.derG}};
    \node[vertex,label=below:$D$](D)[right=of G]    {\summaryRuleSide{#3.derD}};
    \node[vertex,label=below:$H$](H)[right=of D]    {\summaryRuleSide{#3.derH}};
    \draw[normal](D) to (G);
    \draw[normal](D) to (H);
    \draw[normal](L) to (G);
    \draw[normal](K) to (D);
    \draw[normal](R) to (H);
    \end{tikzpicture}
    }
    \end{center}
}

\summaryDerivation{\modIncludeDir/003_r_1463.1010000.0}{\modIncludeDir/003\_r\_1463.1010000.0}{\modIncludeDir/004_r_1463.1010000.0}{\modIncludeDir/004\_r\_1463.1010000.0}
}

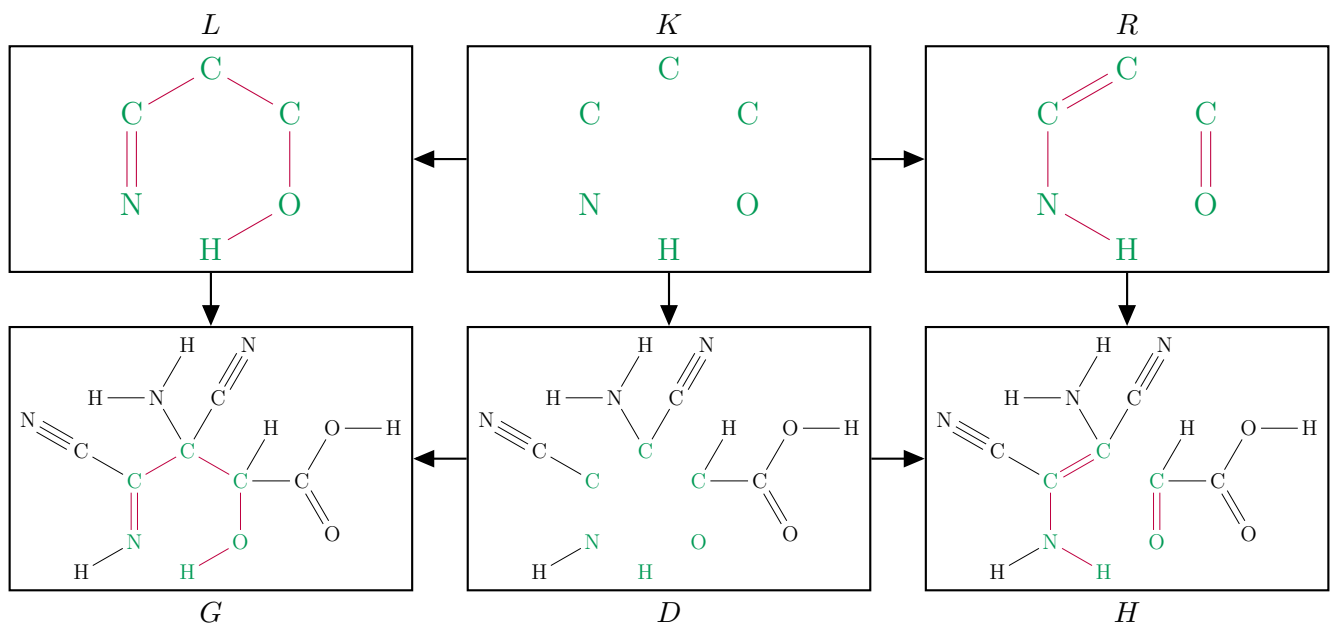
\end{document}